\documentclass[conference]{IEEEtran}
\IEEEoverridecommandlockouts
\usepackage{cite}
\usepackage{multirow}
\usepackage{booktabs}
\usepackage{amsmath,amssymb,amsfonts}
\usepackage{algorithmic}
\usepackage{graphicx}
\usepackage{textcomp}
\usepackage{xcolor}
\usepackage{flushend}
\def\BibTeX{{\rm B\kern-.05em{\sc i\kern-.025em b}\kern-.08em
    T\kern-.1667em\lower.7ex\hbox{E}\kern-.125emX}}
\begin{document}

\title{PF4Microservices: A decomposion scheme for microservices based on Problem Frames\\

 \thanks{*corresponding author: zhili@gxnu.edu.cn}
}

 \author{
 \IEEEauthorblockN{Yitao Bo\IEEEauthorrefmark{2}, Hongbin Xiao\IEEEauthorrefmark{2}, Zhi Li\IEEEauthorrefmark{2}\IEEEauthorrefmark{1}}

 \IEEEauthorblockA{\IEEEauthorrefmark{2}School of Computer Science and Engineering, Guangxi Normal University, Guilin, China} 
%  Email: 2148286607@qq.com,763567198@qq.com, zhili@gxnu.edu.cn\footnote{*corresponding author}
}

\maketitle

\begin{abstract}
In recent years, microservice architecture has become a popular architectural style in software engineering, with its natural support for DevOps and continuous delivery, as well as its scalability and extensibility, which drive industry practitioners to migrate to microservice architecture. However, there are many challenges in adopting a microservice architecture, the most important of which is how to properly decomposition a monolithic system into microservices. Currently, microservice decomposition decisions for monolithic systems rely on subjective human experience, which is a costly, time-consuming process with high uncertainty of results. To address this problem, this paper proposes a method for microservice decomposition using Jackson's Problem Frames. In this method, requirements of the system are analysed, descriptions of the interactions between the proposed software and its environment is obtained, multiple problem diagrams are constructed, and then the problem diagrams are merged by analyzing the correlation and similarity between them, resulting in a microservice decomposition scheme. A case study is also conducted based on a smart parking system. The results of the study show that the method can perform microservice decomposition based on requirements and the software's environment, resulting in reducing the decision-making burden of developers, with reasonable decomposition results.
\end{abstract}

\begin{IEEEkeywords}
Problem Frames, Requirements Engineering, Microservices
\end{IEEEkeywords}

\section{Introduction}

Monolithic Architecture(MA) \cite{b1} is a simple style of software architecture that is easy to develop, test, deploy and extend. As a result, most traditional software systems are implemented using monolithic architectures, where all code is built and deployed in a unified manner and data is managed centrally \cite{b2}. However, as the requirements and the environment change and the software development, the monolithic system will grow in size and complexity and eventually become a huge system. When this happens, the disadvantages of a monolithic system will outweigh its advantages, making bug fixes more difficult, maintenance costs significantly higher and the delivery cycle of new features longer. As a result, over the past few years, with the introduction of Service Oriented Architecture (SOA) \cite{b3} , the rapid development of virtualisation and container technologies, Microservice Architecture \cite{b4} , a style of software architecture that differentiates itself from monolithic architecture, has started to emerge and has become widely accepted and used as a way to overcome the challenges of monolithic architecture\cite{b5}.

In a microservice architecture, a large, complex software system is decomposed into a set of relatively small services that run independently and enable them to communicate with each other through lightweight mechanisms (i.e. RESTful APIs) \cite{b6}. These services are built around business functions and can be developed, tested, deployed and updated independently, which results in shorter delivery cycles for individual services of the system, more flexible technology selection, better scalability, etc. However, there are still many issues that need to be addressed with microservice architectures, the most important of which is how to properly  a monolithic system into microservices \cite{b6}. Most of the current approaches to decomposition monolithic systems into microservices are based on the architect's in-depth knowledge of the business domain, and some scholars have proposed new microservice decomposition methods, most of which are based on source code or historical logs, and few are based on requirements. Most requirements-based microservice decomposition approaches focus on software requirements specifications and do not consider the importance of the software's interaction with the environment.

The Problem Frames (PF) approach \cite{b7}, on the other hand, emphasises the portrayal of the environment in which the system will act, referring the meaning of the requirements to the description of the environment \cite{b8}. Thus, PF is able to capture information about the interactions between the system, requirements and the environment, and a problem diagram is constructed to describe the system, its requirements and the domains that connect them. Therefore, in this paper we choose to use PF for guiding microservice decomposition. The contributions of this paper are summarised as follows.

 \vspace{.1cm}
 \begin{itemize}
     \item This paper proposes a method for microservice decomposition using Jackson's PF. By analysing the requirements of the system, the interaction between the system and the environment, constructing multiple problem diagrams, and analysing the correlation and similarity between the problem diagrams, then the problem diagrams are merged to obtain the microservice decomposition results.
     \item This paper focuses on the basis of microservice decomposition on system requirements and information about the interactions between the system, the requirements and the environment by adopting problem frames approach.
     \item By using PF4Microservices, requirements analysts can design microservices systems based on requirements before developing them, or they can classify microservices based on experience and historical data on the use of monolithic systems when refactoring them into microservices.
 \end{itemize}

The rest of the paper is organised as follows. Section II presents current work on microservice decomposition and the advantages . Section III shows the details of microservice decomposition based on the problem frames. Section IV validates the feasibility of our approach with a real-life case study. Finally, Section V concludes the paper and proposes future work.

\section{Related Work}
As a major challenge for microservice practice, research on microservice decomposing and refactoring has mainly focused on empirical and methodological approaches, and there is still a lack of practical methods and tools to support it \cite{b2}. In this paper, the existing methods are divided into four categories as shown in Table I: software-based requirement specification, software-based design, static source code based, and historical access log based disassembly. In this paper, the approach is to decompose microservices based on the requirements specification of the software.

\textbf{Software-based design approaches}

Hulya et al \cite{b9} propose the design of microservices using domain-driven design (DDD) to obtain the optimal granularity of microservices. Simone et al \cite{b10} model and customise parameters to calculate for coupling cost, communication cost and replication cost to obtain the optimal decomposition model and visualise it. Hassan et al \cite{b11} define adaptable boundaries with architecture elements and uses the Ambient-PRISMA textual language description of the software architecture as input to obtain a microservice decomposition scheme by adjusting the granularity at runtime using workload data.

\textbf{Static source code approaches}

Anfel et al \cite{b12} used a hierarchical clustering algorithm to identify microservices from quality functions and architect recommendations obtained from source code information to get the best decomposition scheme for relevance. Levcovitz et al \cite{b13} performed bottom-up microservice partitioning based on code static dependency diagrams by manually identifying subsystems, classifying database tables. Mazlami et al \cite{b14} obtained the best decomposition scheme for microservices by analysing code and change history, generating a coupling diagram between classes based on the association of code in terms of logic, semantics and importance, and finally clustering the diagram to obtain a microservice decomposition scheme.

\textbf{Historical access logs approaches}

Jin et al \cite{b15} monitored the dynamic behavior of test cases and cluster execution tracking to implement a function-oriented approach to microservice extraction. Ding et al \cite{b2} generated database and code decomposition schemes by dynamically obtaining information about method invocations and database operations during the runtime of the use case in a scenario-driven manner. Mustafa et al \cite{b16} mine through web access logs, session and perform weight assignment, and then use clustering algorithms in order to obtain microservice decomposition schemes. Samuel Lee et al \cite{b17} proposed an automated microservice decomposition and evaluation method DFD-A.

\textbf{Requirement specifications approaches} 

Ahmadvand et al \cite{b18} added use cases to the same microservices based on their scalability, security of aggregated use cases and dependency levels. Chen et al \cite{b6} restructured and compressed the dataflow diagram through business requirement analysis to obtain a decomposable dataflow diagram and then extract microservice candidates. Gysel et al \cite{b19} calculated the correlation between use case graph nodes with 16 coupling criteria and built an undirected weighted graph, using a clustering algorithm to obtain a candidate microservice cut graph. JoonSeok et al. \cite{b20} obtained the business logic diagram by mapping the UML class diagram with a three-level structure and then reconstructed it to obtain the microservice transition diagram. The microservice diagram was obtained by adding invocation APIs based on the ratio of main entities, invocation entity classes, and the ratio of method name similarity.

%%%合并单元格
\begin{table}[htbp]
\caption{Current status of microservice unbundling research}
\label{work}
\centering
\begin{tabular}{cc}		
\toprule 
Types of decomposition methods & Research scholars\\
\midrule  

\multirow{3}*{software-based design} & Hulya 2021  \\
                                   ~ & Simone 2021  \\
                                   ~ & Hassan 2017  \\
                                   \hline
\multirow{3}*{static source code} & Anfel 2020  \\
                                   ~ & Levcovitz 2015  \\
                                   ~ & Mazlami 2017  \\
                                   \hline
\multirow{4}*{historical access logs} & Jin 2018  \\
                                   ~ & Ding Dan 2020  \\
                                   ~ & Mustafa 2017  \\
                                   ~ & Samuel Lee 2021 \\
                                   \hline
\multirow{4}*{requirement specifications} & Ahmadvand 2017  \\
                                   ~ & Chen 2018  \\
                                   ~ & Gysel 2016  \\   
                                   ~ & JookSeok 2020  \\ 
\bottomrule
\end{tabular}
\end{table}

\section{PF4Microservices}
\begin{figure}[htb]
\centerline{\includegraphics[width=3.6in]{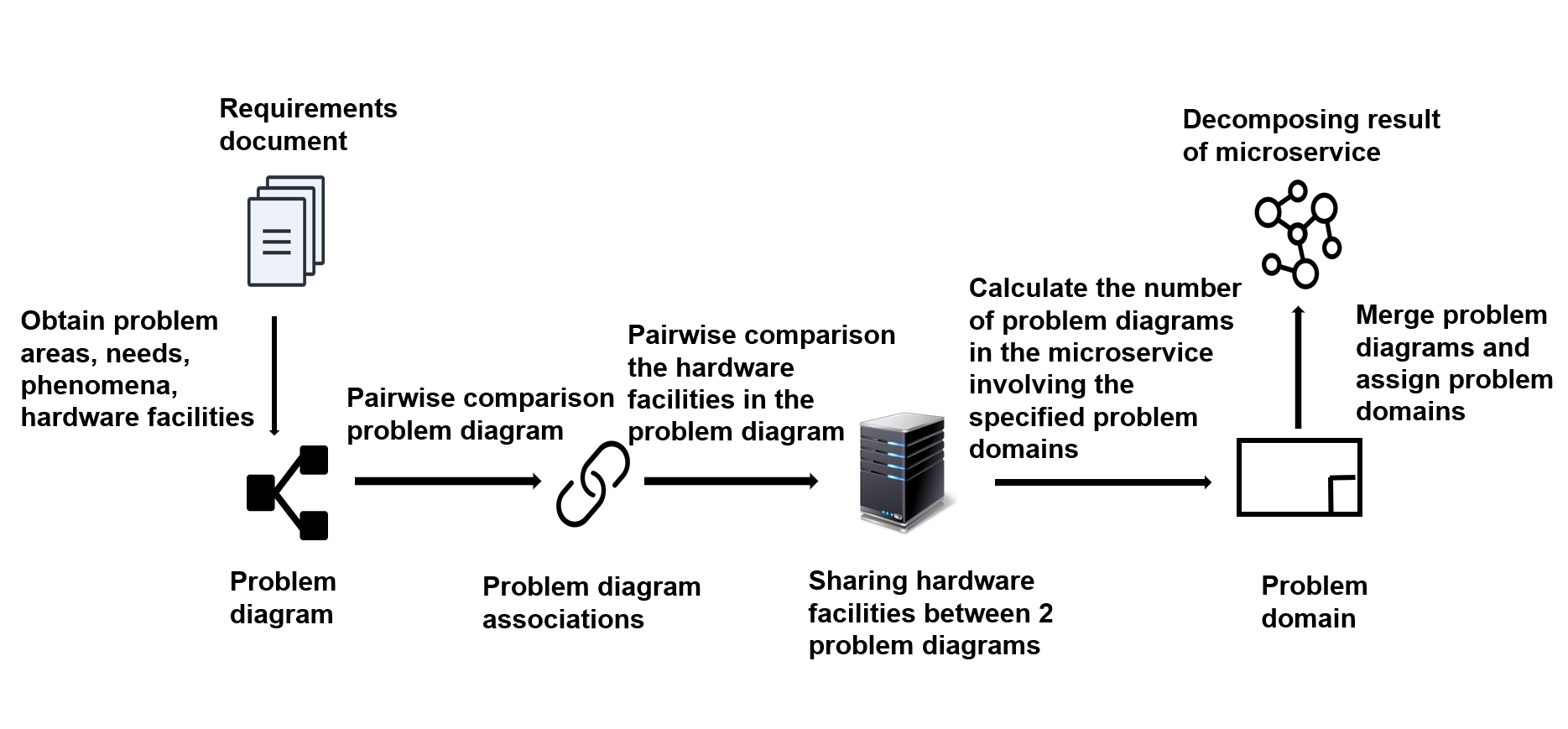}}
\caption{Process of the PF4Microservices method}
\label{weifuwu}
\end{figure}

 The problem frames defines the problem by capturing the characteristics and interconnections of the part of the world it relates to, as well as the concerns and difficulties that may arise [4]. Thus, the problem frames can help focus on the problem rather than getting bogged down in envisioning a solution. The problem frames is able to emphasise not only the world beyond the computer and the desired effect, but also to show the real-world phenomenon and the relationships between phenomenon for the effect achieved. Therefore, in this paper, we have chosen to use the problem frames for microservice decomposition, the process of which is shown in Figure \ref{weifuwu}.

\subsection{Problems Importation}
The requirements document is a content document used to elaborate the product and meet the collaborative staff development. It explains what the product to be developed is, what requirements the product has, what conflicts it has to solve and what purposes it achieves, it not only allows good communication between the requirements analyst and the user, between the requirements analyst and the software designer, but also provides the basis for the confirmation of the software functional requirements. Therefore, this paper takes the requirements document as input, and by analysing the requirements document to obtain the problem domains, requirements, and phenomena in the problem, and by analysing the system topology diagram in the requirements document to obtain the basic hardware facilities and their basic relationships, this paper uses the above information to decompose the microservices.
\subsection{Decomposition Process}

\textbf{STEP1:}After obtaining the basic information from the requirements document, construct a problem diagram by using the problem frames requirements method and add the hardware facilities to the problem diagram after they correspond to the problem domains. In this process, it is important to make the constructed problem diagrams as detailed and realistic as possible.

\textbf{STEP2: }After obtaining multiple problem digrams, the problem digrams are manually compared with each other to obtain the correlation between two problem digrams , so that the problem diagrams with high correlation must be placed in the same microservice, and the problem diagrams with low correlation are judged to be placed in the same microservice according to the actual situation.

\begin{equation}
\mathrm{R}(\mathrm{P_{i}}, \mathrm{P_{j}})=\text { high/low/none }
\end{equation}

\begin{equation}
\mathrm{(\mathrm R(P_{i}, \mathrm P{{j}})=h i g h)\mathrm}\implies M x(P_{i}, P_{j})
\end{equation}      

Where $R(P_{i}, P_{j})$ represents the correlation between two problem diagrams Pi and Pj, and the correlation has three levels: high, low and none, as shown in equation (1). $M_{x}(P_{i}, P_{j})$ represents the placement of problem diagrams $P_{i}$ and $P_{j}$ into microservice $M_{x}$, and when the correlation between problem diagrams is high, they must be placed in the same microservice, as shown in equation (2).

\textbf{STEP3:}The basic hardware facilities used in the problem diagrams are pairwisly compared to get the number of basic hardware facilities shared between the two problem diagrams, and in the process, the number of basic hardware facilities used by all the problem diagrams can also be obtained. When the number of basic hardware facilities used in common between the two problem diagrams is greater than the number of basic hardware facilities used by all the problem diagrams, these two problem diagrams can be placed in the same microservice.

\begin{equation}
(C_{F}(P_{i},P_{j})>C_{F}(All)) \implies M_{x}(P_{i},P_{j})
\end{equation}

Where $C_{F}(All)$ represents the number of basic hardware facilities shared between all problem diagrams and $C_{F}(P_{i},P_{j})$ represents the number of basic hardware facilities shared between two problem diagrams $P_{i}$ and $P_{j}$. When $C_{F}(P_{i},P_{j})$ is greater than $C_{F}(All)$, the problem diagrams $P_{i}$ and $P_{j}$ are placed into the same microservice $M_{x}$, as shown in Equation (3).

\textbf{STEP4:}For problem domains that appear in multiple microservices, the number of problem diagrams involving the specified problem domain in each microservice can be compared to determine which microservice to assign the problem domain to: the problem domain is set to a microservice that has the largest number of problem diagrams involving the specified problem domain in that microservice . If other microservices need to use it, the problem domain is replicated.

\begin{equation}
\begin{split}
\left(C_{P}(M_{x}, D_{k})=\operatorname{Max}\left\{C_{P}(M_{x}, D_{k})\right\}\right)\implies M_{x}(D_{k})
\\
\quad(x=1, \ldots, n ; k=1, \ldots, n)
\end{split}
\end{equation}

Where $C_{P}(M_{x}, D_{k})$ represents the number of problem diagrams in microservice $M_{x}$ involving the specified problem domain $D_{k}$. When $C_{P}(M_{x}, D_{k})$ is the maximum value of the number of problem diagrams in all microservices involving the specified problem domain $D_{k}$, problem domain $D_{k}$ is set to microservice $M_{x}$, as shown in Equation (4).

\subsection{Exporting Microservices}
The output is a microservice decomposition problem diagram: the problem diagrams that should be placed in a microservice are combined to form a problem diagram that contains the problem domains involved and the multiple requirements to be implemented, with one problem diagram representing a microservice module. The diagramical microservice decomposition structure allows requirements analysts, users and software designers to get a clear and quick overview of the end result and to understand the requirements that each microservice implements.

\section{Case Study}
To demonstrate and validate the microservice-oriented decomposition mechanism in this paper using the problem frames requirements analysis approach combined with information on system-environment interactions, we applied our approach to a smart campus system.
\subsection{Background on the Smart Campus System}
Along with the comprehensive interest and deepening construction of smart cities, the wisdom of industrial parks, which are important carriers of urban operations, has come into being. The smart park refers to the development of traditional parks in the domains of infrastructure, resource environment and economic industry, making full use of emerging information technology tools such as the internet, cloud computing and smart technology to intelligently sense, process and coordinate activities and needs within the park, providing a sustainable business environment and an efficient park operation and management environment for enterprises. Therefore, the smart park system needs to provide data transfer and response operations for the interaction between personnel (administrators, employees, etc.), sensor devices (card readers, microwave detectors, etc.) and actuator devices (gate gates, etc.), helping personnel to be able to easily and quickly realise their parking and other features.

\subsection{PF4Microservice Approach in Practice}
\begin{figure}[htb]
\centerline{\includegraphics[width=3.6in]{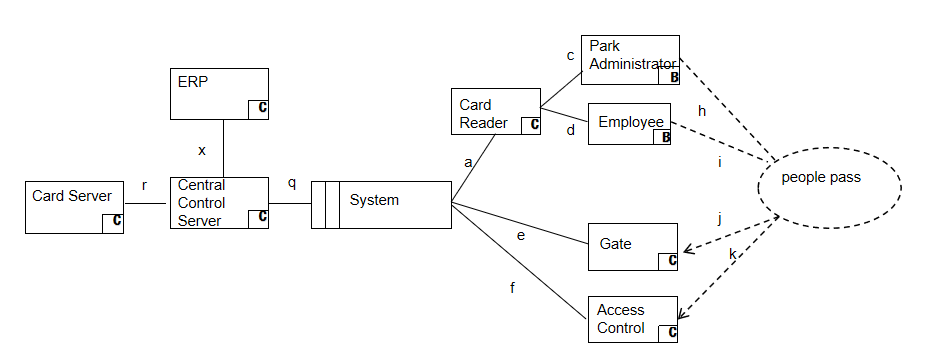}}
\caption{Problem Diagram: Passage of people}
\label{Passage of people}
\end{figure}

\begin{figure}[htb]
\centerline{\includegraphics[width=3.6in]{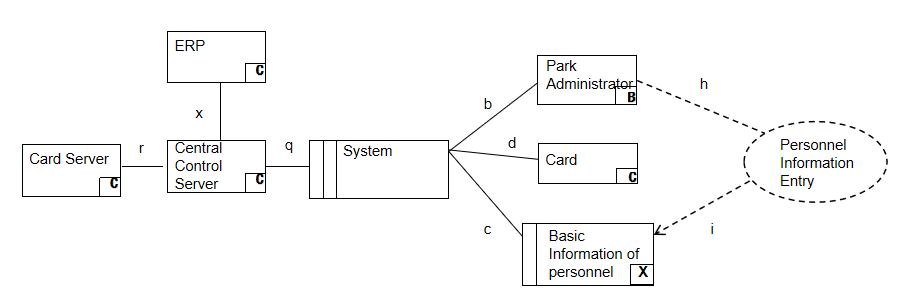}}
\caption{Problem Diagram: Personnel information input}
\label{Personnel information input}
\end{figure}

\begin{figure}[htb]
\centerline{\includegraphics[width=3.6in]{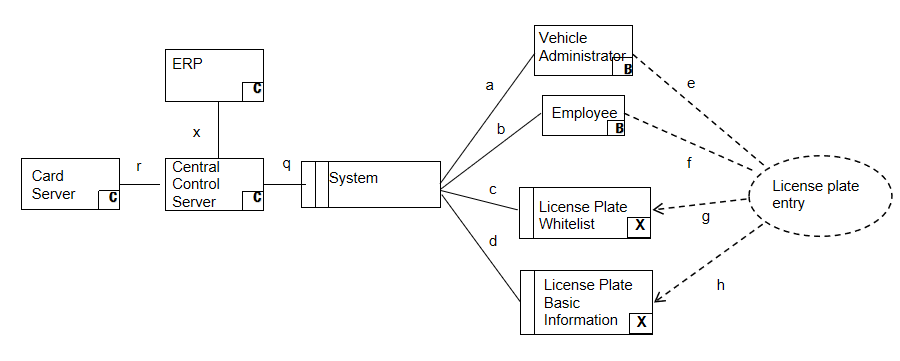}}
\caption{Problem Diagram: License plate entry}
\label{License plate entry}
\end{figure}

\textbf{STEP1:}By analysing the requirements document for the above system, we obtained the 12 requirements and several problem domains and phenomena in the problem. By analysing the system topology diagram, the basic hardware facilities and their basic relationships were obtained. And based on the above information, 12 problem diagrams were constructed, as shown in Table \ref{Requirements}, and the basic hardware facilities were corresponded to the problem domains and added to the problem diagrams. Figure \ref{Passage of people} , Figure \ref{Personnel information input} and Figure \ref{License plate entry}, respectively, show a problem diagram regarding the requirements for personnel access, personnel information entry and vehicle plate entry. Some of the basic elements of the smart campus system regarding security and surveillance requirements, and the need to construct problem diagrams are as follows.

  \vspace{.1cm}
  \begin{itemize}
      \item Machine Domain: System, a rectangle with a double vertical line, indicates that it is a piece of hardware with programmed software running on it.
     
      \item Problem Domain: Administrator, Camera, etc., rectangles without double vertical lines (when the bottom right corner is B, the callable domain, denoted as a person; when the bottom right corner is C, the controlled domain, denoted as a physical device; when the bottom right corner is X, the lexical domain, denoted as a physical description of the data).
     
      \item Requirement: Security Monitor, a dashed oval with arrows linked to the domain, indicating that the requirement is a constraint on the domain, and a dashed line without arrows indicating a dependency.
      
      \item Phenomenon: The solid line indicates the interface between the domains.
      
  \end{itemize}

\begin{table}[htbp]
\caption{Requirements for each microservice}
\centering
\begin{tabular}{|c|c|c|c|}		
\hline \text { Microservices } & \text { Requirements } \\
\hline P1 & \text { Security Monitoring } \\
\hline P2 & \text { Inspection Management } \\
\hline P3 & \text { Intruder Alarm } \\
\hline P4 & \text { Equipment entry } \\
\hline P5 & \text { Top-up with ERP embedded APP } \\
\hline P6 & \text { Top up with the Campus General APP } \\
\hline P7 & \text { On-site top-up } \\
\hline P8 & \text { Access for people } \\
\hline P9 & \text { Personnel directory } \\
\hline P10 & \text { License plate entry } \\
\hline P11 & \text { Vehicle entry } \\
\hline P12 & \text { Vehicle exit } \\
\hline

\end{tabular}
\label{Requirements}
\end{table}

\textbf{STEP2:}
By comparing the problem diagrams pairwisly, the correlation between the problem diagrams is obtained, as shown in Table \ref{corre}. It is obtained that problem diagrams P2 and P3 should be placed in the same microservice, and problem diagrams P11 and P12 should be placed in the same microservice.

\begin{table}[htbp]
\caption{Correlation between problem diagrams}
\begin{tabular}{|p{0.3cm}|p{0.2cm}|p{0.2cm}|p{0.3cm}|p{0.2cm}|p{0.2cm}|p{0.2cm}|p{0.2cm}|p{0.2cm}|p{0.2cm}|p{0.3cm}|p{0.3cm}|p{0.3cm}|}		
\hline             & P1 &  P2 & P3 & P4 & P5 & P6 & P7 & P8 & P9 & P10 & P11 & P12 \\
\hline P1          & /  &  /  & /  &low& / & / & / & / & / & / & / &/ \\                         
\hline P2          &    &  /  &\textcolor[rgb]{1,0,0}{high}&low& / & / & / & / & / & / & / &/ \\                           
\hline P3          &    &     & /  &low& / & / & / & / & / & / & / &/ \\ 
\hline P4          &    &     &    & / & / & / & / & / & / & / & / &/ \\
\hline P5          &    &     &    &   &/  & / & / & / & / & / & / &/ \\
\hline P6          &    &     &    &   &   & / & / & / & / & / & / &/ \\
\hline P7          &    &     &    &   &   &   & / & / & / & / & / &/ \\ 
\hline P8          &    &     &    &   &   &   &   & / &low& / & / &/ \\ 
\hline P9          &    &     &    &   &   &   &   &   & / &low& / &/ \\
\hline P10         &    &     &    &   &   &   &   &   &   & / &low&low\\                   
\hline P11         &    &     &    &   &   &   &   &   &   &   & / &\textcolor[rgb]{1,0,0}{high}\\                    
\hline P12         &    &     &    &   &   &   &   &   &   &   &   &/ \\                     
\hline 

\end{tabular}
\label{corre}
\end{table}

\textbf{STEP3:}
The basic hardware facilities used in the problem diagrams are compared two by two to obtain the number of basic hardware facilities used in common between all problem diagrams, as shown in Table \ref{Number}, and the number of basic hardware facilities used by all problem diagrams can be obtained as 2. Therefore, when the number of basic hardware facilities used in common between problem diagrams is greater than 2, they can be placed in the same microservice. The microservice decomposition structure is obtained: M1 (P1, P2, P3, P4), M2 (P5, P6, P7), M3 (P8, P9, P10), M4 (P11, P12). Since problem diagrams with low relevance are judged to be placed in the same microservice based on the actual situation, and problem diagrams P10 and P11, P12 are closely related in the actual situation and share more problem domains, this paper decides to merge microservices M3 and M4 into one microservice to obtain the microservice decomposition structure: M1 (P1, P2, P3, P4), M2 (P5, P6, P7), M3 (P8, P9, P10, P11, P12).

\begin{table}[htbp]
\caption{Number of basic hardware facilities in common use between problem diagrams}
\begin{tabular}{|p{0.3cm}|p{0.2cm}|p{0.2cm}|p{0.2cm}|p{0.2cm}|p{0.2cm}|p{0.2cm}|p{0.2cm}|p{0.2cm}|p{0.2cm}|p{0.3cm}|p{0.3cm}|p{0.3cm}|}		
\hline             & P1 &  P2 & P3 & P4 & P5 & P6 & P7 & P8 & P9 & P10 & P11 & P12 \\
\hline P1          & /  &  \textcolor[rgb]{1,0,0}{3}  & \textcolor[rgb]{1,0,0}{3}  & \textcolor[rgb]{1,0,0}{3} & 2 & 2 & 2 & 2 & 2 & 2 & 2 &2 \\                         
\hline P2          &    &  /  & \textcolor[rgb]{1,0,0}{4}  & \textcolor[rgb]{1,0,0}{3} & 2 & 2 & 2 & 2 & 2 & 2 & 2 &2 \\                           
\hline P3          &    &     & /  & \textcolor[rgb]{1,0,0}{3} & 2 & 2 & 2 & 2 & 2 & 2 & 2 &2 \\ 
\hline P4          &    &     &    & / & 2 & 2 & 2 & 2 & 2 & 2 & 2 &2 \\
\hline P5          &    &     &    &   &/  & \textcolor[rgb]{1,0,0}{5} & \textcolor[rgb]{1,0,0}{5} & 2 & 2 & 2 & 2 &2 \\
\hline P6          &    &     &    &   &   & / & \textcolor[rgb]{1,0,0}{5} & 2 & 2 & 2 & 2 &2 \\
\hline P7          &    &     &    &   &   &   & / & 2 & 2 & 2 & 2 &2 \\ 
\hline P8          &    &     &    &   &   &   &   & / & \textcolor[rgb]{1,0,0}{3} & 2 & 2 &2 \\ 
\hline P9          &    &     &    &   &   &   &   &   & / & 2 & 2 &2 \\
\hline P10         &    &     &    &   &   &   &   &   &   & / &\textcolor[rgb]{1,0,0}{3} &\textcolor[rgb]{1,0,0}{3} \\                   
\hline P11         &    &     &    &   &   &   &   &   &   &   & / &\textcolor[rgb]{1,0,0}{3} \\                    
\hline P12         &    &     &    &   &   &   &   &   &   &   &   &/ \\                     
\hline 

\end{tabular}
\label{Number}
\end{table}

\textbf{STEP4:}
For problem domains that appear in multiple microservices: Camera, Employee, ERP, Central Control Server, the problem diagrams involving the specified problem domains in each microservice are obtained, as shown in Table \ref{DIA}. Thus, Camera is assigned to M3, Employee to M3, ERP to M3, Central Control Server to M3, and the problem domains are replicated if other microservices need to use them.

\begin{table}[htbp]
\caption{Problem diagram for each microservice involving the specified problem domain}
\begin{tabular}{|c|c|c|c|}		
\hline \text { Problem Domain } & \text { M1 }  &  M2 &  M3\\
\hline Camera &  P1 & / & P11, P12 \\
\hline Employee & /  & P5, P7  &  P8, P10, P12   \\
\hline ERP & P1, P2, P3, P4 & P5, P6, P7 & P8, P9, P10, P11, P12  \\
\hline Central Control & P1, P2, P3, P4 & P5, P6, P7  & P8, P9, P10, P11, P12  \\

\hline

\end{tabular}
\label{DIA}
\end{table}

\subsection{Generated microservice architecture}
\begin{figure}[htb]
\centerline{\includegraphics[width=3.6in]{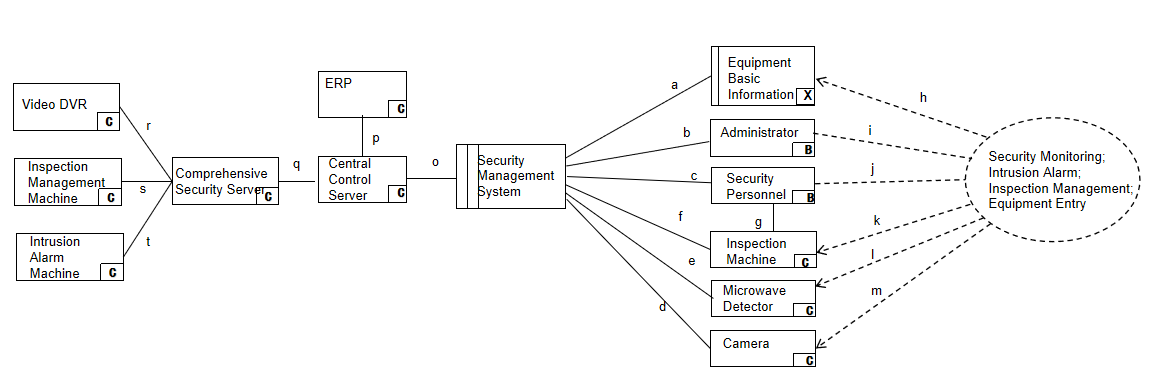}}
\caption{Problem Diagram: Security Management}
\label{Security Management}
\end{figure}

\begin{figure}[htb]
\centerline{\includegraphics[width=3.6in]{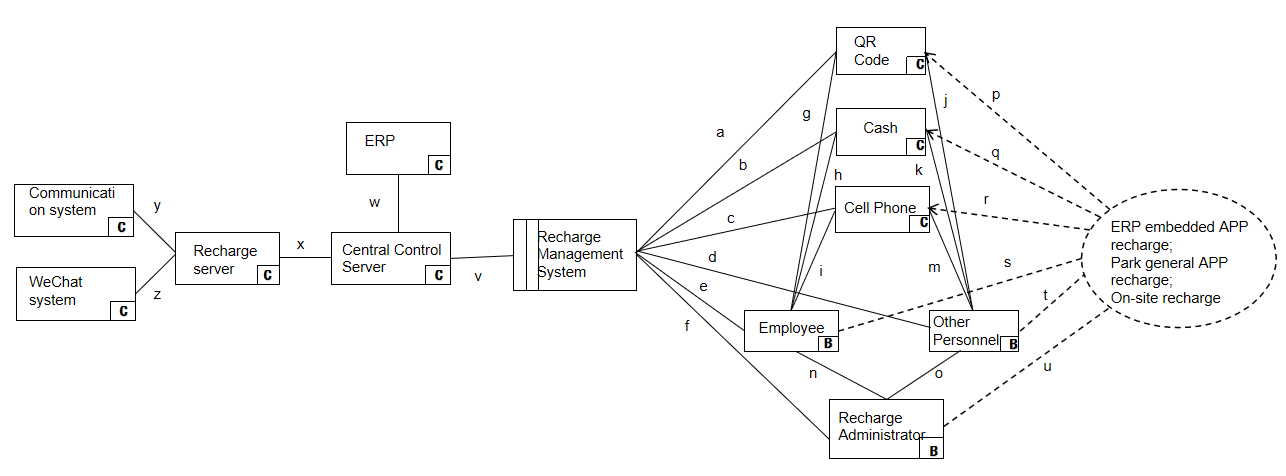}}
\caption{Problem Diagram: Recharge Management}
\label{Recharge Management}
\end{figure}

\begin{figure}[htb]
\centerline{\includegraphics[width=3.6in]{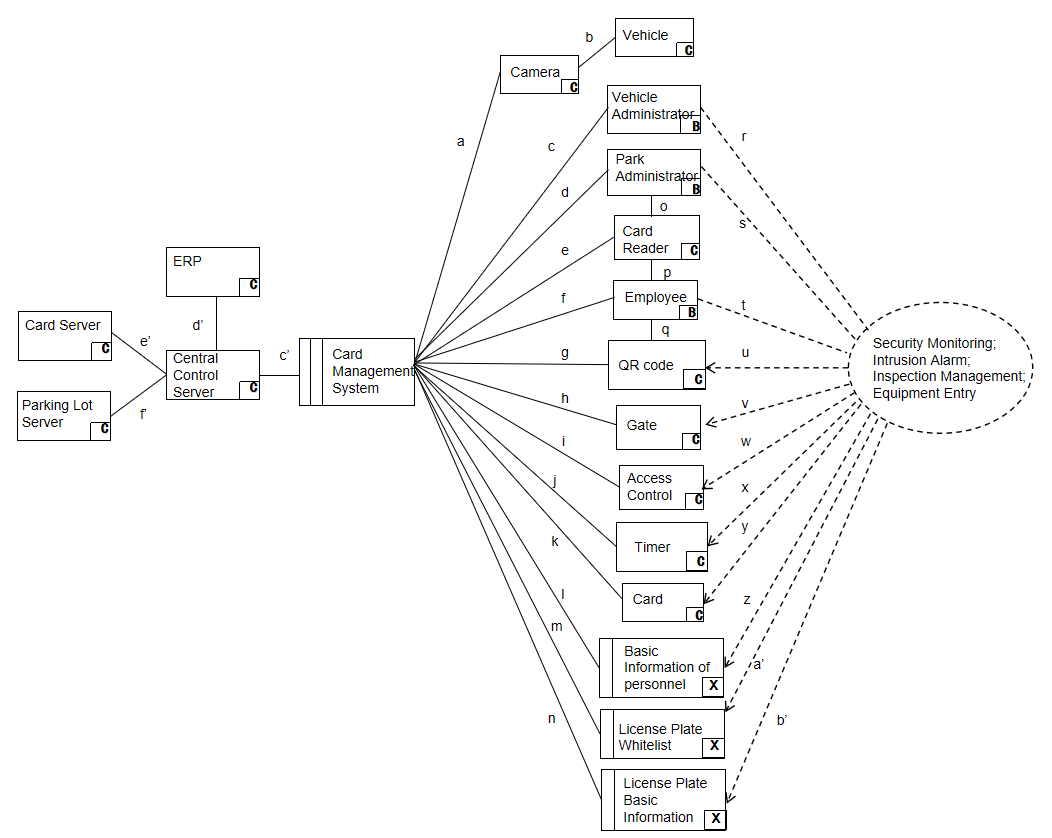}}
\caption{Problem Diagram: One-Card Management}
\label{One-Card Management}
\end{figure}

\begin{figure}[htb]
\centerline{\includegraphics[width=3.6in]{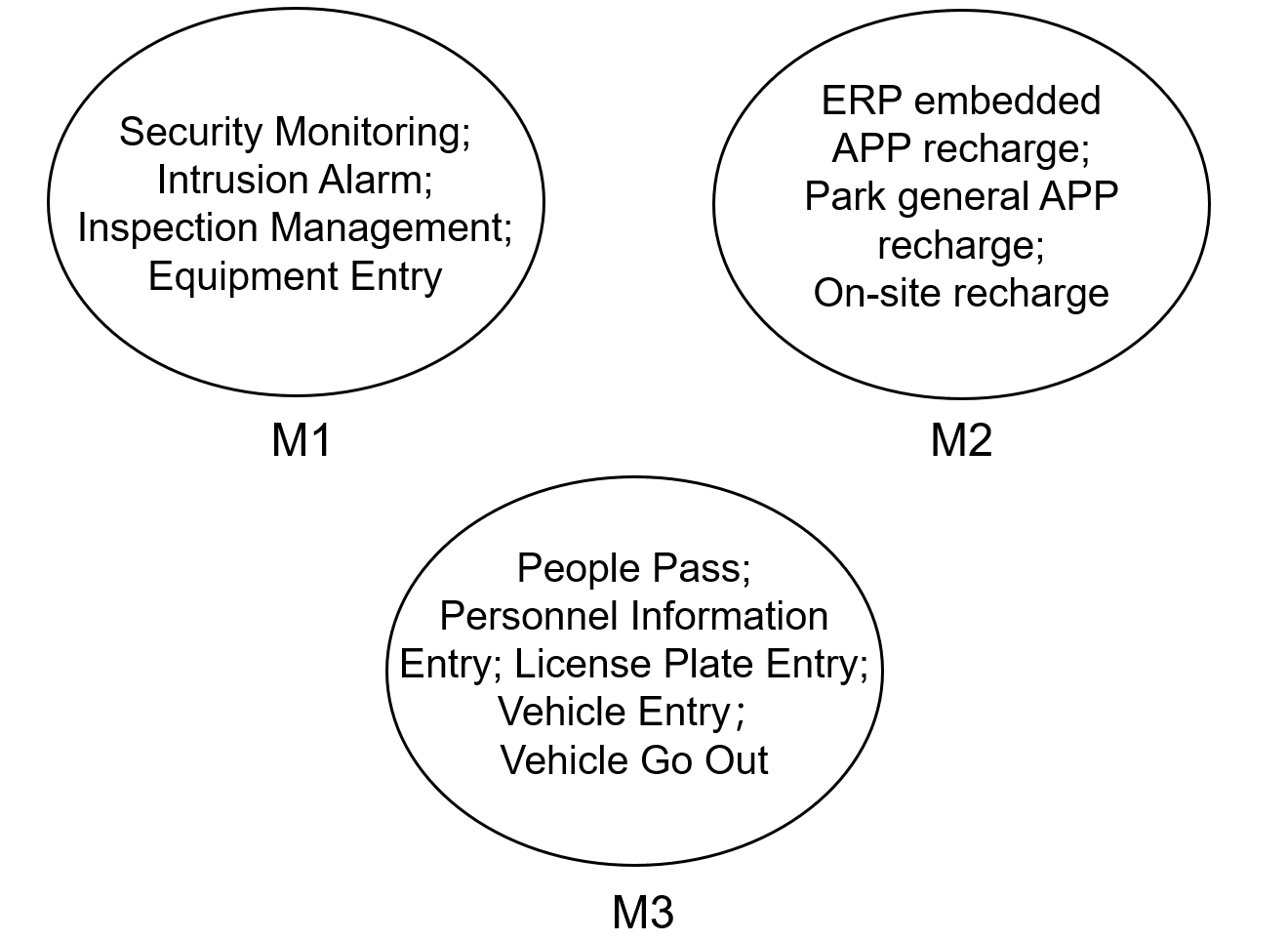}}
\caption{Microservice architecture based on problem diagram}
\label{Microservice architecture}
\end{figure}

The problem diagrams assigned to the same microservice are eventually merged to obtain a problem diagram containing multiple requirements, and a microservice can be built based on one problem diagram. Therefore, the final three microservice architectures obtained are based on: the security management problem graph, the top-up management problem graph and the one-card management problem graph, as shown in Figure \ref{Security Management}, Figure \ref{Recharge Management} and Figure \ref{One-Card Management}. And the microservice architecture obtained based on the problem diagram is shown in Figure \ref{Microservice architecture}, where the requirements to be implemented are shown in the microservice.

We claim that the decomposition results using PF4Microservices are reasonable because they are similar to the results in \cite{b18}.

\subsection{Methodological evaluation}
The case study demonstrates the advantages of our PF4Microservices research on microservices decomposition in terms of software-based requirements specification.

\textbf{Advantage 1:}
Our approach is applicable in cases where the relevant artefacts are not well developed during the architectural design phase of the software system.

\textbf{Advantage 2:}
Our approach enables the observation of the system side of the operation and the data interaction in terms of the environment, capturing the interaction between the system and the environment.

\textbf{Advantage 3:}
Our approach uses a rigorous and detailed requirements specification. By using this approach, the microservice splitter has a general understanding of the overall structure of the system and its business processes, on the basis of which multiple problem diagrams can be constructed to obtain the domains that needs to be covered to implement each requirement, which can then be decomposed to obtain the microservices decompositio result.

\section{Conclusion And Future Work}
In order to address the challenges of migrating monolithic architecture to microservice architecture, this paper proposes a method for microservice disaggregation using the problem frames. The microservice-oriented decomposition process can be divided into three stages: firstly, multiple problem diagrams are constructed by analysing the requirements of the system and obtaining information on the interactions between the system, requirements and the environment. Secondly, the problem diagrams are merged by analysing the correlation between them and the similarity of the basic hardware facilities. Finally, problem domains that appear in multiple microservices can be assigned to which microservice by comparing the number of problem diagrams involving the specified problem domain in each microservice, and the microservice decomposition result is obtained. The experiments validate the feasibility of the approach and the correctness of the decomposition results, and show that the problem frames captures information about the interactions between the system, requirements and the environment and displays this information in problem diagrams, giving the requirements analyst a quick and clear systematic view of the system to be decomposed and the microservice decomposition to meet the desired requirements, taking into account the environment.

Considering that our current work still has corresponding limitations in terms of the time spent on microservice  decomposition, we will develop our future work along three directions. Firstly, we will propose methods to automatically extract problem areas, requirements and phenomena from requirements documents, reducing the time required to construct problem diagrams. Secondly, we will consider using existing correlation calculation methods to obtain accurate correlations between problem diagrams, and then decomposing them according to the degree of correlation between the refined problem diagrams. Thirdly, a prototype microservice decomposing will be implemented to enable semi-automation of microservice decomposition.

\section{Acknowledgements}
This work is partially supported by the National Natural Science Foundation of China (61862009), Guangxi Natural Science Foundation (2012GXNSFCA053010), Guangxi "Bagui Scholar” Teams for Innovation and Research, the Project of the Guangxi Key Lab of Multi-source Information Mining \& Security (Director’s grant 19-A-01-02), Guangxi Collaborative Innovation Center of Multi-source Information Integration and Intelligent Processing.


\begin{thebibliography}{00}
\bibitem{b1} C. Richardson, "Pattern: Monolithic architecture," 22 June 2017. [Online]. Available: http://microservices.io/patterns/monolithic.html

\bibitem{b2} Ding, D., Peng, X., Guo, X. F., Zhang, J., \& Wu, Y. J. (2020). Scenario-driven and Bottom-up Microservice Decomposition Method for Monolithic Systems. Ruan Jian Xue Bao/Journal of Software, 31(11), 3461–3480. https://doi.org/10.13328/j.cnki.jos.006031

\bibitem{b3} M. P. Papazoglou, “Service-oriented computing: Concepts, characteristics and directions,” in Web Information Systems Engineering, 2003.WISE 2003. Proceedings of the Fourth International Conference on.IEEE, 2003, pp. 3–12.

\bibitem{b4} Lewis J, Fowler M. Microservices: A definition of this new architectural term. 2014. https://martinfowler.com/articles/microservices.html

\bibitem{b5} Zhou X, Peng X, Xie T, et al. Fault analysis and debugging of microservice systems: Industrial survey, benchmark system, and empirical study[J]. IEEE Transactions on Software Engineering, 2018, 47(2): 243-260.

\bibitem{b6} Chen, R., Li, S., \& Li, Z. (2018). From Monolith to Microservices: A Dataflow-Driven Approach. Proceedings - Asia-Pacific Software Engineering Conference, APSEC, 2017-December, 466–475. https://doi.org/10.1109/APSEC.2017.53

\bibitem{b7} Jackson M (2001) Problem Frames Analyzing and Structuring Software Development Problems. Addison-Wesley

\bibitem{b8}  Z.Jin, Xiaohong Chen, Z. Li, and Yijun Yu. RE4CPS: Requirements engineering for cyber-physical systems. In Proceedings of the 27th IEEE International Conference on Requirements Engineering (Vol. 2019-September, pp.496–497).IEEE Computer Society,Jeju Island, South Korea. https://doi.org/10.1109/RE.2019.00072

\bibitem{b9} Vural, H., \& Koyuncu, M. (2021). Does Domain-Driven Design Lead to Finding the Optimal Modularity of a Microservice? IEEE Access, 9, 32721–32733. https://doi.org/10.1109/ACCESS.2021.3060895

\bibitem{b10} Staffa, S., Quattrocchi, G., Margara, A., \& Cugola, G. (2021). Pangaea: Semi-automated Monolith Decomposition into Microservices (Vol. 825480, Issue 825480). Springer International Publishing. https://doi.org/10.1007/978-3-030-91431-8-60

\bibitem{b11} Hassan, S., Ali, N., \& Bahsoon, R. (2017). Microservice Ambients: An Architectural Meta-Modelling Approach for Microservice Granularity. Proceedings - 2017 IEEE International Conference on Software Architecture, ICSA 2017, 1–10. https://doi.org/10.1109/ICSA.2017.32

\bibitem{b12} Selmadji, A., Seriai, A. D., Bouziane, H. L., Oumarou Mahamane, R., Zaragoza, P., \& Dony, C. (2020). From monolithic architecture style to microservice one based on a semi-automatic approach. Proceedings - IEEE 17th International Conference on Software Architecture, ICSA 2020, Section III, 157–168. https://doi.org/10.1109/ICSA47634.2020.00023

\bibitem{b13} Levcovitz, A., Terra, R., \& Valente, M. T. (2016). Towards a Technique for Extracting Microservices from Monolithic Enterprise Systems. http://arxiv.org/abs/1605.03175

\bibitem{b14} Mazlami, G., Cito, J., \& Leitner, P. (2017). Extraction of Microservices from Monolithic Software Architectures. Proceedings - 2017 IEEE 24th International Conference on Web Services, ICWS 2017, 524–531. https://doi.org/10.1109/ICWS.2017.61

\bibitem{b15} Jin, W., Liu, T., Zheng, Q., Cui, D., \& Cai, Y. (2018). Functionality-Oriented Microservice Extraction Based on Execution Trace Clustering. Proceedings - 2018 IEEE International Conference on Web Services, ICWS 2018 - Part of the 2018 IEEE World Congress on Services, October, 211–218. https://doi.org/10.1109/ICWS.2018.00034 

\bibitem{b16} Mustafa, O., \& J. Marx Gómez. (2017). Optimizing economics of microservices by planning for granularity level - Experience Report. 2017 Programming Technology for the Future Web (ProWeb), September, 6.

\bibitem{b17} Li, S. S., Rong, G. P., Gao, Q. Y., \& Shao, D. (2021). Optimized Dataflow-driven Approach for Microservices-oriented Decomposition. Ruan Jian Xue Bao/Journal of Software, 32(5), 1284–1301. https://doi.org/10.13328/j.cnki.jos.006233

\bibitem{b18} Ahmadvand, M., \& Ibrahim, A. (2017). Requirements reconciliation for scalable and secure microservice (de)composition. Proceedings - 2016 IEEE 24th International Requirements Engineering Conference Workshops, REW 2016, September, 68–73. https://doi.org/10.1109/REW.2016.14

\bibitem{b19} Gysel, M., Kölbener, L., Giersche, W., \& Zimmermann, O. (2016). Service cutter: A systematic approach to service decomposition. Lecture Notes in Computer Science (Including Subseries Lecture Notes in Artificial Intelligence and Lecture Notes in Bioinformatics), 9846 LNCS, 185–200. https://doi.org/10.1007/978-3-319-44482-6-12

\bibitem{b20} Park, J., Kim, D., \& Yeom, K. (2020). An Approach for Reconstructing Applications to Develop Container-Based Microservices. Mobile Information Systems, 2020. https://doi.org/10.1155/2020/4295937

\end{thebibliography}
\end{document}